# Physica Status Solidi B: Basic Solid State Physics
## Amorphous p-type conducting Zn-xIr oxide (x>0.13) thin films deposited by reactive magnetron co-sputtering
### --Manuscript Draft--

| | |
|---|---|
| **Manuscript Number:** | pssb.202100374R1 |
| **Article Type:** | Research Article |
| **Corresponding Author:** | Juris Purans, Prof.<br>Institute of Solid State Physics RAS: Institut fiziki tverdogo tela Rossijskoj akademii nauk<br>Riga, LATVIA |
| **Corresponding Author E-Mail:** | purans@cfi.lu.lv |
| **Order of Authors:** | Juris Purans, Prof. |
| | Martins Zubkins, DR. |
| | Janis Timoshenko |
| | Jevgenijs Gabrusenoks, Dr. |
| | Kaspars Pudzs, Dr. |
| | Andris Azens, Dr. |
| | Qin Wang, Dr. |
| **Keywords:** | zinc-iridium oxide; amorphous thin film; reactive magnetron co-sputtering; RMC-EXAFS; p-type conductivity |
| **Section/Category:** | CSW-2021 - Compound Semiconductors |
| **Abstract:** | Zinc-Iridium Oxide (Zn-Ir-O) thin films have been demonstrated as a $p$-type conducting material. However, the stability of $p$-type conductivity with respect to chemical composition or temperature is still unclear. This study focuses on the local atomic structure and the electrical properties of Zn-Ir-O films in the large Ir concentration range. The films are deposited by reactive DC magnetron co-sputtering at two different substrate temperatures – without intentional heating and 300 °C. EXAFS reveals that strongly disordered $ZnO_4$ tetrahedra are the main Zn complexes in Zn-Ir-O films up to 67.4 at.% Ir. As the Ir concentration increases, an effective increase of Ir oxidation state is observed. Reverse Monte Carlo analysis of EXAFS at Zn K-edge shows that the average Zn-O interatomic distance and disorder factor increase with the Ir concentration. We observed that the nano-crystalline $w$-ZnO structure is preserved in a wider Ir concentration range if the substrate is heated during deposition. At low Ir concentration, the transition from $n$- to $p$-type conductivity is observed regardless of the substrates temperature. Electrical resistivity decreases exponentially with the Ir concentration in the Zn-Ir-O films, and it is slightly lower in the case of the heated films. |
| **Author Comments:** | We are aware that some of the results and conclusions in this manuscript overlap with previously published data. However, the reason for the overlap is to provide a comprehensive study, and most results in this article are not published elsewhere:<br>•the results for samples deposited at 300 ° C (XRD, Raman, electrical conductivity and thermoelectric measurements). The thermoelectric measurements of these samples suggest that the p-type conductivity in Zn-Ir-O films is thermally stable. This is explained in more detail in the revised manuscript. In addition, it is shown that the p-type conductivity is maintained over a wide range of Ir concentrations regardless of the deposition temperature (only low Ir concentrations in paper: Thin Solid Films 2017);<br>•reverse Monte Carlo (RMC) approach has not previously been used to analyze EXAFS spectra at Zn K edge of Zn-Ir-O films. The RMC analysis clearly shows the local atomic structure around Zn ions and gives quantitative results on the increased structural disorder upon an increase in Ir concentration. For the first time, the XANES and EXAFS spectra at Zn K edge are shown over a wide Ir concentration range. In paper Mater. Chem. C 9 2021, 4948, only the study of the local atomic structure around Ir ions in ZnO thin films with different iridium content is conducted. |

|  | We believe that the results on the local atomic structure and the electrical measurements of Zn-Ir-O films (up to 70 at.% Ir) deposited at evelated temperatures provide a reasonable contribution to a better understanding of the physics behind the p-type conductivity. In addition, to the best of our knowledge, the Hall effect measurements of pure $IrO_2$ thin films performed in our study, cannot be found elsewhere. The explanation of the novelty has been improved in the revised manuscript. |
|---|---|
| **Additional Information:** | |
| **Question** | **Response** |
| Please submit a plain text version of your cover letter here. | Dear editor, we wish to submit a revised manuscript entitl" Amorphous p-type conducting Zn-xIr oxide (x>0.13) thin films deposited by reactive magnetron co-sputtering" for consideration by the Thin Solid Films. The authors are grateful to both reviewers and editor for useful comments that made it possible to improve the text of the article. We have made corrections in the text following the reviewers comments. All corrections throughout the manuscript are marked in yellow. The manuscript has been professionally proofread by PRS (Proof-reading-service).

We confirm that this work is original and has not been published elsewhere nor is it currently under consideration for publication elsewhere.

In this paper, for the first time we report on the structure and the optical and electrical properties of ZnO:Ir thin films in the full range of Zn-Ir concentrations as well at different deposition temperatures. This is significant because it is unknown whether it is possible to achieve p-type conductivity at different Ir contents and deposition temperatures, and how the type and the magnitude of conductivity are affected by the film structure. The paper should be of interest to readers in the areas of transparent conductive oxides and thin film deposition.
Although doped ZnO thin films are promising n-type TCO materials, obtaining p-type ZnO thin films would be an important milestone in transparent electronics, allowing the production of wide band gap p-n homo-junctions.
Please address all correspondence concerning this manuscript to me at purans@cfi.lu.lv.

Thank you for your consideration of this manuscript.
Sincerely,
Dr. Prof. Juris Purans |
| Do you or any of your co-authors have a conflict of interest to declare? | No. The authors declare no conflict of interest. |
| This journal's Expects Data Policy requires a Data Availability Statement (even if no data are shared), which will be published alongside your manuscript if it is accepted for publication.

Do you choose to share the research data described in this manuscript? | No. Research data are not shared. |

| Response to Reviewers: | Response Letter |
|---|---|
| | The authors are grateful to both reviewers and editor for useful comments that made it possible to improve the text of the article. We have made corrections in the text following the reviewers comments. All corrections throughout the manuscript are marked in yellow.<br><br>COMMENTS TO AUTHOR:<br><br>Reviewer #1: The manuscript reports on the synthesis through reactive magnetron co-sputtering of amorphous p-type conducting Zn-xIr oxide (x>0.13) thin films and their characterization. The topic of the manuscript is certainly of interest for the PSS b readership.<br>In particular the authors report the effect of a wide range of Ir% doping into ZnO,in regard to structural and electrical properties. The manuscript contents are reasonable, but my main concern regards the similarity of results to authors previously published papers.<br>The effect of lower doping was already addressed in ref.7, paper titled "Changes in structure and conduction type upon addition of Ir to ZnO thin films" and published by the authors in Thin Solid Film lower in 2017.<br>The effect of the higher doping was addressed in ref.14 titled "Raman, electron microscopy and electrical transport studies of x-ray amorphous Zn-Ir-O thin films deposited by reactive DC magnetron sputtering", published in 2015. Other studies (Xanes and EXAFS) are reported in J. Mater. Chem. C 2021.<br>So my concern is what is the novelty of this manuscript? It is not only a problem of the identical nature of deposited films, but also of type of characterization carried out that is the same than that previously reported.<br>Response:<br>We are aware that some of the results and conclusions in this manuscript overlap with previously published data. However, the reason for the overlap is to provide a comprehensive study, and most results in this article are not published elsewhere:<br>•the results for samples deposited at 300 ° C (XRD, Raman, electrical conductivity and thermoelectric measurements). The thermoelectric measurements of these samples suggest that the p-type conductivity in Zn-Ir-O films is thermally stable. This is explained in more detail in the revised manuscript. In addition, it is shown that the p-type conductivity is maintained over a wide range of Ir concentrations regardless of the deposition temperature (only low Ir concentrations in paper: Thin Solid Films 2017);<br>•reverse Monte Carlo (RMC) approach has not previously been used to analyze EXAFS spectra at Zn K edge of Zn-Ir-O films. The RMC analysis clearly shows the local atomic structure around Zn ions and gives quantitative results on the increased structural disorder upon an increase in Ir concentration. For the first time, the XANES and EXAFS spectra at Zn K edge are shown over a wide Ir concentration range. In paper Mater. Chem. C 9 2021, 4948, only the study of the local atomic structure around Ir ions in ZnO thin films with different iridium content is conducted.<br><br>We believe that the results on the local atomic structure and the electrical measurements of Zn-Ir-O films (up to 70 at.% Ir) deposited at evelated temperatures provide a reasonable contribution to a better understanding of the physics behind the p-type conductivity. In addition, to the best of our knowledge, the Hall effect measurements of pure IrO2 thin films performed in our study, cannot be found elsewhere. The explanation of the novelty has been improved in the revised manuscript.<br><br>The authors should also carefully check the language, which needs to be carefully revised.<br>Response:<br>The language is carefully revised by the authors.<br><br>Overall, in my opinion, the present manuscript cannot be published in the present form and it should be significantly modified/improved to make clear what are the differences with respect to the previously published papers.<br><br>Reviewer #2: This is a nice study of Ir-doped ZnO thin films. The properties of the thin films are explored as a function of Ir content, with an emphasis on the conductivity |

properties of this material, as this system has been considered as a possible candidate for p-type transparent conductors. I have two suggestions to make:

- In the description of the RMC procedure, I would like a bit of discussion about why a 4x4x4 ZnO cell was selected, and no other size.
Response:
To select the size of the supercell for RMC simulations, we relied on our previous experience with RMC simulations for bulk ZnO and other oxides, see Ref[25-26]. There it was found that at least ca. 100 absorbing atoms are needed in the structure model to ensure proper averaging of calculated EXAFS spectra. Moreover, periodic boundary conditions introduce artificial correlations in the atomic displacements, if the supercell is too small (i.e., is not larger than the largest interatomic distance that contributes to the calculated EXAFS spectrum). On the other hand, using too large supercell sizes does not improve the accuracy of the structure model (due to the fact that EXAFS is local method and normally does not contain information on atoms that are more than few angstroms away from the absorbing atom), but significantly increases the computational costs of RMC simulations, since both the convergence of RMC method is slowed down, and the CPU time required for each iteration is increased. 4x4x4 supercell provides a reasonable compromise that works well for most simple oxide materials. We have added this information to manuscript text.

- In page 9, line 32 "The is reflected...", I did not understand the meaning of this sentence.
Response:
The sentence corrected "This is also reflected … ".

Reviewer #3: I have accepted to review the submitted manuscript pssb.202100374. However looking precisely on the paper, it appears that the abstract does not really describe this paper.
Response:
The abstract has been improved by adding XAS RMC results as main results, and the novelty is better explained.

In fact, the main part of the paper is devoted to the XAS measurements. The structural characteristics and resistivity properties only represent a small part of the manuscript. Concerning these rather limited results on the physical properties of the films, I dont think that this small part brings a lot of new results in the field of Zn-Ir-O films.
Response:
We believe that the quantitative results on the local atomic structure around Zn ions provided by RMC-EXAFS analysis gives an important information about the structural units in amorphous Zn-Ir-O films. The RMC approach is more suitable for strongly disordered structures compared to conventional EXAFS fitting. Thermoelectric measurements show that p-type conductivity is maintained over a wide range of Ir concentrations, and deposition at elevated temperatures indicates the themal stability of the p-type conductivity.

Dr.habil.phys. Juris Purāns
Institute of Solid State Physics, University of Latvia
Kengaraga 8, LV-1063, Riga, Latvia

29.07.2021

Dear editor,

we wish to submit a new manuscript entitl" Amorphous p-type conducting Zn-xIr oxide (x>0.13) thin films deposited by reactive magnetron co-sputtering" for consideration by the Thin Solid Films. We confirm that this work is original and has not been published elsewhere nor is it currently under consideration for publication elsewhere. The manuscript has been professionally proofread by PRS (Proof-reading-service).

In this paper, for the first time we report on the structure and the optical and electrical properties of ZnO:Ir thin films in the full range of Zn-Ir concentrations as well at different deposition temperatures. This is significant because it is unknown whether it is possible to achieve $p$-type conductivity at different Ir contents and deposition temperatures, and how the type and the magnitude of conductivity are affected by the film structure. The paper should be of interest to readers in the areas of transparent conductive oxides and thin film deposition.

Although doped ZnO thin films are promising $n$-type TCO materials, obtaining $p$-type ZnO thin films would be an important milestone in transparent electronics, allowing the production of wide band gap $p$-$n$ homo-junctions.

Please address all correspondence concerning this manuscript to me at purans@cfi.lu.lv.

Thank you for your consideration of this manuscript.

Sincerely,

Dr. Prof. Juris Purans

# Response Letter

The authors are grateful to both reviewers and editor for useful comments that made it possible to improve the text of the article. We have made corrections in the text following the reviewers comments. <u>All corrections throughout the manuscript are marked in yellow.</u>

COMMENTS TO AUTHOR:

Reviewer #1: The manuscript reports on the synthesis through reactive magnetron co-sputtering of amorphous p-type conducting Zn-xIr oxide (x>0.13) thin films and their characterization. The topic of the manuscript is certainly of interest for the PSS b readership.

In particular the authors report the effect of a wide range of Ir% doping into ZnO, in regard to structural and electrical properties. The manuscript contents are reasonable, but my main concern regards the similarity of results to authors previously published papers.

The effect of lower doping was already addressed in ref.7, paper titled "Changes in structure and conduction type upon addition of Ir to ZnO thin films" and published by the authors in Thin Solid Film lower in 2017.

The effect of the higher doping was addressed in ref.14 titled "Raman, electron microscopy and electrical transport studies of x-ray amorphous Zn-Ir-O thin films deposited by reactive DC magnetron sputtering", published in 2015. Other studies (Xanes and EXAFS) are reported in J. Mater. Chem. C 2021.

So my concern is what is the novelty of this manuscript? It is not only a problem of the identical nature of deposited films, but also of type of characterization carried out that is the same than that previously reported.

***Response:***

*We are aware that some of the results and conclusions in this manuscript overlap with previously published data. However, the reason for the overlap is to provide a comprehensive study, and most results in this article are not published elsewhere:*

- *the results for samples deposited at 300 ° C (XRD, Raman, electrical conductivity and thermoelectric measurements). The thermoelectric measurements of these samples suggest that the p-type conductivity in Zn-Ir-O films is thermally stable. This is explained in more detail in the revised manuscript. In addition, it is shown that the p-type conductivity is maintained over a wide range of Ir concentrations regardless of the deposition temperature (only low Ir concentrations in paper: Thin Solid Films 2017);*
- *reverse Monte Carlo (RMC) approach has not previously been used to analyze EXAFS spectra at Zn K edge of Zn-Ir-O films. The RMC analysis clearly shows the local atomic structure around Zn ions and gives quantitative results on the increased structural disorder upon an increase in Ir concentration. For the first time, the XANES and EXAFS spectra at Zn K edge are shown over a wide Ir concentration range. In paper Mater. Chem. C 9 2021, 4948, only the study of the local atomic structure around Ir ions in ZnO thin films with different iridium content is conducted.*

*We believe that the results on the local atomic structure and the electrical measurements of Zn-Ir-O films (up to 70 at.% Ir) deposited at evelated temperatures provide a reasonable contribution to a better understanding of the physics behind the p-type conductivity. In addition, to the best of our knowledge, the Hall effect measurements of pure $IrO_2$ thin films performed in our study, cannot be found elsewhere. The explanation of the novelty has been improved in the revised manuscript.*

The authors should also carefully check the language, which needs to be carefully revised.

*Response:*

*The language is carefully revised by the authors.*

Overall, in my opinion, the present manuscript cannot be published in the present form and it should be significantly modified/improved to make clear what are the differences with respect to the previously published papers.

Reviewer #2: This is a nice study of Ir-doped ZnO thin films. The properties of the thin films are explored as a function of Ir content, with an emphasis on the conductivity properties of this material, as this system has been considered as a possible candidate for p-type transparent conductors. I have two suggestions to make:

- In the description of the RMC procedure, I would like a bit of discussion about why a 4x4x4 ZnO cell was selected, and no other size.

*Response:*

*To select the size of the supercell for RMC simulations, we relied on our previous experience with RMC simulations for bulk ZnO and other oxides, see Ref[25-26]. There it was found that at least ca. 100 absorbing atoms are needed in the structure model to ensure proper averaging of calculated EXAFS spectra. Moreover, periodic boundary conditions introduce artificial correlations in the atomic displacements, if the supercell is too small (i.e., is not larger than the largest interatomic distance that contributes to the calculated EXAFS spectrum). On the other hand, using too large supercell sizes does not improve the accuracy of the structure model (due to the fact that EXAFS is local method and normally does not contain information on atoms that are more than few angstroms away from the absorbing atom), but significantly increases the computational costs of RMC simulations, since both the convergence of RMC method is slowed down, and the CPU time required for each iteration is increased. 4x4x4 supercell provides a reasonable compromise that works well for most simple oxide materials. We have added this information to manuscript text.*

- In page 9, line 32 "The is reflected...", I did not understand the meaning of this sentence.

*Response:*

*The sentence corrected "This is also reflected … ".*

Reviewer #3: I have accepted to review the submitted manuscript pssb.202100374. However looking precisely on the paper, it appears that the abstract does not really describe this paper.

*Response:*

*The abstract has been improved by adding XAS RMC results as main results, and the novelty is better explained.*

In fact, the main part of the paper is devoted to the XAS measurements. The structural characteristics and resistivity properties only represent a small part of the manuscript. Concerning these rather limited results on the physical properties of the films, I dont think that this small part brings a lot of new results in the field of Zn-Ir-O films.

*Response:*

*We believe that the quantitative results on the local atomic structure around Zn ions provided by RMC-EXAFS analysis gives an important information about the structural units in amorphous Zn-Ir-O films. The RMC approach is more suitable for strongly disordered structures compared to conventional EXAFS fitting. Thermoelectric measurements show that p-type conductivity is maintained over a wide range of Ir concentrations, and deposition at elevated temperatures indicates the themal stability of the p-type conductivity.*

# Amorphous *p*-type conducting Zn-*x*Ir oxide (*x*>0.13) thin films deposited by reactive magnetron co-sputtering

*Martins Zubkins, Janis Timoshenko, Jevgenijs Gabrusenoks, Kaspars Pudzs, Andris Azens, Qin Wang, Juris Purans\**

Dr. M. Zubkins, Dr. J. Timoshenko, Dr. J. Gabrusenoks, Dr. K. Pudzs, Dr. A. Azens, Dr. J. Purans
Institute of Solid State Physics, University of Latvia, Kengaraga 8, LV-1063 Riga, Latvia
E-mail: purans@cfi.lu.lv

Dr. Q. Wang
*Department of Smart Hardware, Reasearch Institutes of Sweden, box 1070, 164 25 Stockholm, Sweden*

Zinc-Iridium Oxide (Zn-Ir-O) thin films have been demonstrated as a *p*-type conducting material. However, the stability of *p*-type conductivity with respect to chemical composition or temperature is still unclear. This study focuses on the local atomic structure and the electrical properties of Zn-Ir-O films in the large Ir concentration range. The films are deposited by reactive DC magnetron co-sputtering at two different substrate temperatures – without intentional heating and 300 °C. EXAFS reveals that strongly disordered ZnO$_4$ tetrahedra are the main Zn complexes in Zn-Ir-O films up to 67.4 at.% Ir. As the Ir concentration increases, an effective increase of Ir oxidation state is observed. Reverse Monte Carlo analysis of EXAFS at Zn K-edge shows that the average Zn-O interatomic distance and disorder factor increase with the Ir concentration. We observed that the nano-crystalline *w*-ZnO structure is preserved in a wider Ir concentration range if the substrate is heated during deposition. At low Ir concentration, the transition from *n*- to *p*-type conductivity is observed regardless of the substrates temperature. Electrical resistivity decreases exponentially with the Ir concentration in the Zn-Ir-O films, and it is slightly lower in the case of the heated films..

## 1. Introduction

WILEY-VCH

To develop transparent electronics, it is still essential to search for reliable transparent *p*-type semiconductors. Kawazoe et al.[1] reported the *p*-type conductivity in the highly transparent thin film of copper aluminium oxide. Since this publication, a variety of *p*-type transparent conductive oxides (*p*-TCOs) have been investigated. Despite these results, their insufficient transparency and low conductivity continue restricting the technological applications of the *p*-TCOs.[2-3] A possible solution is to rely on precious metal-based transparent oxides. Several of such materials have been shown to exhibit the *p*-type conductivity, including polycrystalline Zn$M_2$O$_4$ (*M*= Ir, Rh. Co) films.[4-6]

It has been shown that the conductivity type of zinc-iridium oxide (Zn-Ir-O) films changes from *n*- to *p*-type by increasing an Ir concentration from 12 to 16 at.%.[7] However, changes in the local structure of Zn upon an increase in Ir concentration have not been evaluated. In addition, the stability of *p*-type conductivity to changes in composition or in high temperatures has not been tested. According to DFT calculations performed in Ref.[8], IrO$_x$ ($x$ = 4,5,6) complexes show *p*-type conductivity with the Fermi level up to 0.8 eV from the valence band maximum (VBM) if the Ir concentration is greater than 12.5%. The theoretical calculations of the substitution defects of Ir$^{2+}_{Zn}$ and Ir$^{3+}_{Zn}$ in Ir-doped ZnO predict localized energy states in the band gap which would reduce the transmittance of ZnO:Ir films.[9,10] The decrease of visible light transmittance in Zn-Ir-O with Ir concentration has been observed experimentally.[11,12] The structural analysis of doped IrO$_2$ electrocatalysts with the general composition of Ir$_{1-x}$$M_x$O$_2$ (*M* = Co, Ni and Zn, 0.05≤x≤0.2) prepared by a hydrolysis method shows that the doping elements enter the lattice positions in rutile structure of iridium dioxide.[13] Analysis of the local structure of the catalysts based on EXAFS shows that the dopant cations are not homogeneously distributed but have a tendency to form clusters.

Zn-Ir-O has hardly ever been systematically investigated in a wide Ir concentration range. Therefore, in this study we employ XRD, XAS and Raman techniques to investigate the structure of Zn-Ir-O films (up to ≈ 70 at.% Ir) deposited by reactive DC magnetron co-sputtering.

Reverse Monte Carlo (RMC) simulations of the Zn K-edge EXAFS spectra are performed to study the local structure of Zn ions. The electrical conductivity as well as the thermoelectric measurements are also presented to determine the electrical conductivity of the films and to observe the transition from *n*-type to *p*-type conductivity. To evaluate the structure of films deposited at higher temperatures and the thermal stability of *p*-type conductivity, the films are also deposited at 300 °C.

## 2. Experimental details

Zn-Ir-O as well as pure $ZnO_x$ and $IrO_{2-x}$ thin films were deposited on soda-lime glass, Ti and polyimide type substrates by reactive DC magnetron co-sputtering in an Ar (20 sccm) + $O_2$ (10 sccm) atmosphere (10 mTorr working pressure). Two types of sputtering methods were used to deposit the studied films. Films of the first type (mainly with high Ir concentration) were deposited by sputtering a metallic Zn (99.95 wt%) target with Ir (99.6 wt%) pieces on the target erosion zone. Ir concentration in the films was varied by changing Ir amount on the Zn target erosion zone. Samples of the second type (mainly with low Ir concentration) were deposited by sputtering metallic Zn (99.95 wt%) and Ir (99.6 wt%) targets simultaneously. The power on the Ir target was used as a composition control parameter. Detailed experimental procedures of both methods can be found in Ref. [7] and [14]. Elemental analysis of the films was carried out by an X-ray fluorescence spectrometer (XRF), Eagle III. Because with XRF it is difficult to quantify elements lighter than sodium, our measurements show only the Ir to Zn atomic concentration ratio. To determine the influence of the substrate temperature on the film's structure and properties, two sets of samples were deposited: one set without intentional substrate heating during the deposition and the second one with the additional heating at a temperature of 300 °C. All the studied samples together with the deposition parameters are summarized in **Table 1**.

**Table 1.** Deposition parameters, thickness and Ir/Zn atomic concentration ratio of the studied Zn-Ir-O films on glass, Ti, and polyimide substrates.

| Sample | Sputtering target(s) | Sputtering power (W) | Ir area on the Zn target erosion zone (%) | Thickness (nm) | Substrate temperature (°C) | Ir/Zn atomic concentration ratio (%) |
| --- | --- | --- | --- | --- | --- | --- |
| $ZnO_x$ | Zn | 200 | - | 388 | Not heated | 0.0 |
| Zn-Ir-O | Zn and Ir | 200 (Zn), 6 (Ir) | - | 566 | Not heated | 1.7 |
| Zn-Ir-O | Zn and Ir | 200 (Zn), 10 (Ir) | - | 581 | Not heated | 3.0 |
| Zn-Ir-O | Zn and Ir | 200 (Zn), 20 (Ir) | - | 659 | Not heated | 5.1 |
| Zn-Ir-O | Ir pieces on Zn | 100 | ≈0.7 | 393 | Not heated | 7.0 |
| Zn-Ir-O | Zn and Ir | 200 (Zn), 40 (Ir) | - | 752 | Not heated | 12.4 |
| Zn-Ir-O | Zn and Ir | 200 (Zn), 70 (Ir) | - | 778 | Not heated | 16.1 |
| Zn-Ir-O | Ir pieces on Zn | 100 | ≈3.0 | 308 | Not heated | 29.4 |
| Zn-Ir-O | Ir pieces on Zn | 100 | ≈5.8 | 244 | Not heated | 33.6 |
| Zn-Ir-O | Ir pieces on Zn | 100 | ≈9.6 | 141 | Not heated | 44.6 |
| Zn-Ir-O | Ir pieces on Zn | 100 | ≈12.2 | 121 | Not heated | 53.5 |
| Zn-Ir-O | Ir pieces on Zn | 100 | ≈14.6 | 95 | Not heated | 67.4 |
| $IrO_{2-x}$ | Ir | 100 | - | 108 | Not heated | 100.0 |
| $ZnO_x$ | Zn | 200 | - | 388 | 300 | 0.0 |
| Zn-Ir-O | Zn and Ir | 200 (Zn), 6 (Ir) | - | 165 | 300 | 2.3 |
| Zn-Ir-O | Zn and Ir | 200 (Zn), 10 (Ir) | - | 164 | 300 | 3.2 |
| Zn-Ir-O | Zn and Ir | 200 (Zn), 15 (Ir) | - | 193 | 300 | 5.6 |
| Zn-Ir-O | Zn and Ir | 200 (Zn), 25 (Ir) | - | 233 | 300 | 8.0 |
| Zn-Ir-O | Zn and Ir | 200 (Zn), 40 (Ir) | - | 190 | 300 | 13.8 |
| Zn-Ir-O | Zn and Ir | 200 (Zn), 70 (Ir) | - | 282 | 300 | 19.6 |
| Zn-Ir-O | Zn and Ir | 200 (Zn), 90 (Ir) | - | 226 | 300 | 24.1 |
| Zn-Ir-O | Zn and Ir | 200 (Zn), 110 (Ir) | - | 286 | 300 | 33.0 |
| Zn-Ir-O | Zn and Ir | 200 (Zn), 130 (Ir) | - | 281 | 300 | 36.9 |
| Zn-Ir-O | Ir pieces on Zn | 100 | ≈14.6 | 185 | 300 | 61.5 |
| $IrO_{2-x}$ | Ir | 100 | - | 176 | 300 | 100.0 |

The XRD measurements were done on a PANalytical X'Pert PRO diffractometer equipped with the Cu anode X-ray tube and a multichannel solid-state detector. The Zn K-edge (9659 eV) and Ir $L_3$-edge (11215 eV) X-ray absorption spectra (XAS) were measured in transmission mode at the SOLEIL synchrotron bending-magnet beamline Samba [15] at ambient conditions. More details of the experiment can be found in Ref. [7]. Raman spectroscopy measurements were performed at room temperature by a SPEX1403 monochromator with multichannel detectors and an inVia Renishaw Raman microscope. Both an Ar laser (514.5 nm) and YAG second harmonics laser (532 nm) were used as the excitation source. The electrical properties of the films were investigated by Hall effect measurement system HMS5000 at room temperature. To determine the sign of the Seebeck coefficient, the thermoelectric measurements in plane were performed by a self-assembled measurement system. More details of the system can be found in Ref.[7].

## 3. Results and discussion

### 3.1. XRD measurements

The X-ray diffractograms of the Zn-Ir-O films are shown in **Figure 1**. It can be clearly seen that the crystallinity of the films deteriorates when Ir concentration is increased. Pure $ZnO_x$ films contain nano-crystallites with the structure of wurtzite type ZnO (*w*-ZnO). This structure is observed even for the not-heated samples due to the rapid crystallisation of ZnO at room temperature. An X-ray amorphous structure of ZnO can be obtained if a deposition is performed at cryogenic temperatures.[16] The nano-crystallites have a preferred (002) orientation in the direction of *c*-axis. The preferred orientation decreases with the Ir concentration and an additional (101) maximum appears. At this point, both (002) and (101) diffraction maximums are shifted toward lower angles compared to the *w*-ZnO reference X-ray diffractogram (PDF card No.: 01-070-8072). The further increase of the Ir concentration changes the structure from nano-crystalline to X-ray amorphous. The nano-crystalline structure is preserved in the larger Ir concentration range if the substrate temperature is 300 ºC during the deposition. It has been previously shown that the Zn-Ir-O films become completely amorphous in the Ir concentration range from 7 to 16 at.% if the substrates are not heated intentionally.[7] The X-ray amorphous structure is preserved up to the pure $IrO_{2-x}$ film for the non-heated samples. In turn, two shifted diffraction maxima, which correspond to the (110) and (200) planes of rutile $IrO_2$ (*r*-$IrO_2$) structure (PDF card No.: 00-015-0870), appear for the pure IrO$_{2-x}$ film that was deposited on the heated substrate.

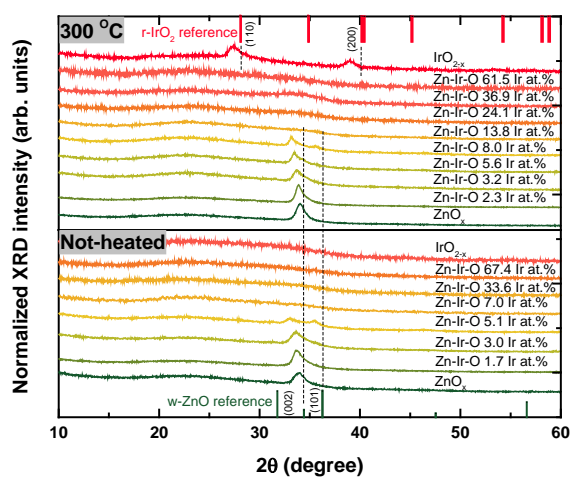

**Figure 1.** X-ray diffractograms of the Zn-Ir-O films at different Ir concentrations both for heated (300 °C) and non-heated samples during the deposition.

### 3.2. XAS measurements

X-ray absorption spectroscopy (XAS) is one of the most important techniques for investigations of the local atomic structure and oxidation state of the atom of interest in a broad range of materials, including the X-ray amorphous structures. In this work, we have analysed both the X-ray absorption near-edge structure (XANES) and extended X-ray absorption fine structure (EXAFS) parts of XAS spectra, collected at Ir L$_3$-edge and Zn K-edge. Here XANES spectra, extending up to ca. 40 eV above the absorption edge, are very sensitive to the charge density (i.e., the evolution of the oxidation state) and bonding motifs around the absorbing atom. EXAFS spectra, in turn, extending from ca. 40 eV up to 1 keV above the absorption edge, are important for the determination of neighbouring atoms distribution.[17-20]

Zn K-edge XANES spectra (**Figure 2(a)**) for all Zn-Ir-O samples align well with the reference spectrum for bulk *w*-ZnO, suggesting 2+ oxidation state for Zn in Zn-Ir-O films, and the presence of ZnO$_4$ tetrahedra as the main structural units. However, XANES spectra for our thin films appear relatively featureless with respect to that of bulk ZnO, suggesting a strongly disordered environment for Zn species. The changes in the relative intensity of the main Zn K-edge XANES features upon an increase in Ir concentration reflect the different degree of distortion of $ZnO_4$ tetrahedra.[21]

Collected Ir $L_3$-edge XANES spectra (**Figure 2(b)**) for Zn-Ir-O films show a systematic shift of the main feature toward higher energies upon increased Ir concentration, which can be linked to the changes in the density of available vacant states in the *d*-orbital for the excited $2p_{1/2}$ electrons, and suggests an effective increase of Ir oxidation state.[19, 22, 23]

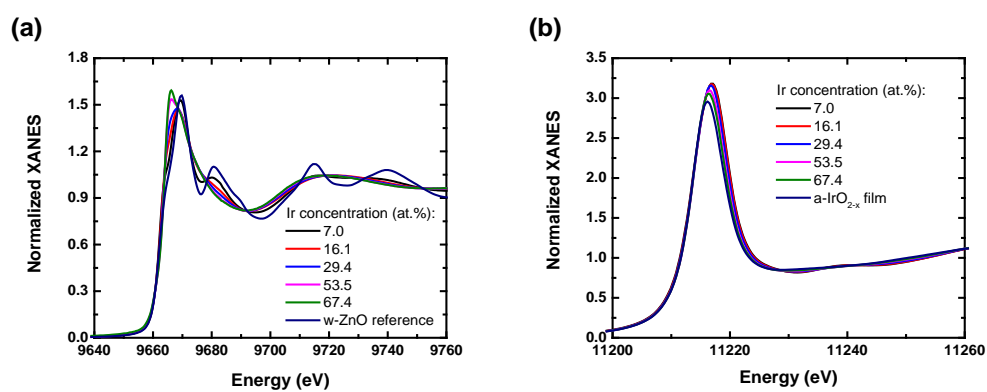

**Figure 2.** Zn K-edge (a) un Ir $L_3$-edge (b) XANES spectra of the Zn-Ir-O films with different Ir atomic concentrations. XANES spectra of reference compounds – polycrystalline bulk *w*-ZnO taken from Ref. [24] and the amorphous (*a*-) $IrO_{2-x}$ film deposited in this study – are also shown.

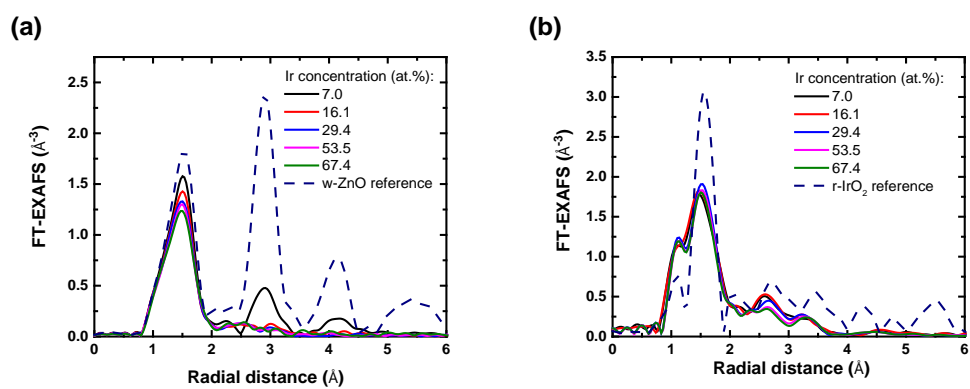

**Figure 3.** Moduli of the FT-EXAFS spectra at Zn K-edge (a) and Ir $L_3$-edge (b) for the Zn-Ir-O films and reference compounds – polycrystalline bulk *w*-ZnO taken from Ref. [24] and *r*-$IrO_2$

taken from Ref. [20]. (Note: distances in the FT-EXAFS do not correspond to the real distances due to the phase shifts of the signal.)

Further insight into the local structure of Zn and Ir species is provided by the analysis of EXAFS spectra, shown in **Figures 3(a)** and **3(b)**, respectively. In the case of Zn K-edge EXAFS, Fourier-transformed (FT) spectra are dominated by the contribution of first coordination shell (Zn-O bonds). Similar position and the intensity of the main FT-EXAFS peak in all Zn-Ir-O samples confirm that 4-coordinated Zn species with Zn-O bond length similar to that in *w*-ZnO are the main structural units in our thin films, in agreement with XANES results. The slight decrease in the intensity of FT-EXAFS peaks upon an increase in Ir concentration suggests an increase in structural disorder (Debye-Waller factors).

Strong structural disorder results also in suppression of more distant FT-EXAFS peaks. Figure 3(a) highlights the fundamental difference between the samples with higher Ir content and that with 7.0 at.% Ir. Only in the sample with the lowest Ir concentration we observed presence of significant 2nd and 3rd peaks in FT-EXAFS, which resemble those in *w*-ZnO, suggesting that wurtzite-type structure is locally preserved in this sample. No such peaks were observed for Zn-Ir-O peaks with higher Ir loading, suggesting their amorphous nature. We emphasize here also the lack of significant Zn-Ir bond contributions in our EXAFS data.

To fit the contributions of distant coordination shell, we employ reverse Monte Carlo (RMC) method, as described in Ref. [25] and implemented within EvAX code.[26] Here EXAFS spectra are fitted in an iterative stochastic process, where we start with ideal *w*-ZnO structure, and introduce small random displacements in the positions of all atoms within 4×4×4 ZnO supercell with periodic boundary conditions, until a good agreement is obtained between experimental Zn K-edge EXAFS spectrum and simulated spectrum for the structure model. The size of the supercell was chosen based on our previous experience with RMC simulations for bulk ZnO and other oxides, see Ref[25-26], and is chosen to be large enough to ensure proper averaging of calculated EXAFS spectra and avoid artifacts due to periodic boundary conditions employed, but, at the same time, to ensure reasonably fast convergence of RMC method. For EXAFS spectra calculations we use FEFF8.5 code [27] and include multiple-scattering effects with up to 7 scattering events. See Ref. [25] for more details. The advantage of RMC approach in comparison to conventional EXAFS fitting is that it allows fitting of contributions of distant coordination shells, but also, crucially, allows fitting of EXAFS spectra for strongly distorted structures, where the conventional EXAFS fitting, which commonly relies on the assumption that the bond length distribution has near-Gaussian shape, results in significant systematic errors.[28]

The results of RMC fitting for Zn K-edge EXAFS spectra are shown in **Figure 4(a)** and **4(b)**. One needs to emphasize here that due to limited information available from EXAFS spectra of our disordered thin films, the obtained structure model is not unambiguous (especially for Ir-rich thin films, where only the first coordination shell contribute significantly to the experimental data). Nevertheless, the good agreement between simulated and experimental EXAFS spectra confirms that the $ZnO_4$ tetrahedral units are the main Zn species in all the samples, and we expect also that our simulations provide reliable information on the shape of the radial distribution function (RDF) at least for the first coordination shell (**Figure 4(c)**).

The obtained RDFs show increasingly asymmetric and broadened shape for Zn-O RDF upon an increase in the Ir concentration. This is also reflected in the increased average Zn-O interatomic distance and increased disorder factor, calculated from the atomic coordinates in the final RMC model (**Table 2**).

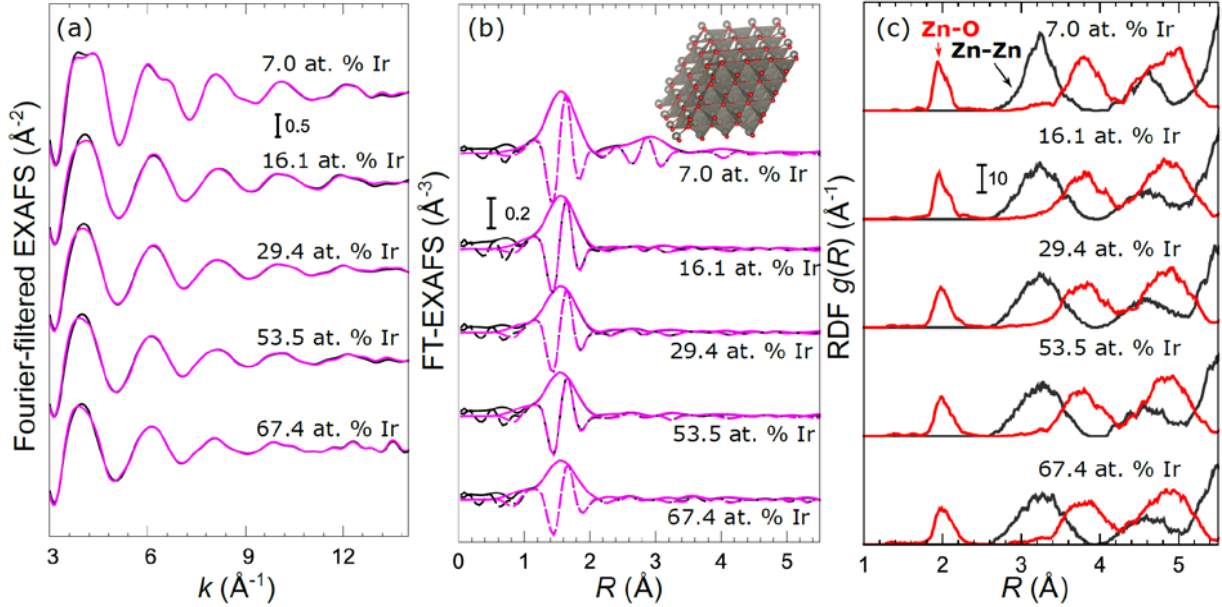

**Figure 4.** RMC fitting results for Zn K-edge EXAFS for Zn-Ir-O films with different Ir content. Comparison of experimental spectra with the ones calculated for the final RMC model in *k*-space after Fourier-filtering (a) and in *R*-space after Fourier transformation (b). The final structure model for the sample with 7.0 at.% Ir is shown in the inset. Partial RDFs for Zn-O and Zn-Zn bonds (c). Spectra and RDFs are shifted vertically for clarity.

**Table 2.** Average Zn-O interatomic distances and corresponding disorder factor (Debye-Waller factor $\sigma^2$) for the Zn-Ir-O films as obtained from RMC-EXAFS analysis and Zn K-edge. Uncertainties of the last digit are given in parentheses.

| Ir concentration (at.%) | $R$ (Å)  | $\sigma^2$ (Å$^2$) |
|---|---|---|
| 7.0  | 1.991(1) | 0.0081(1) |
| 16.1 | 1.995(1) | 0.0099(2) |
| 29.4 | 1.995(1) | 0.0119(2) |
| 53.5 | 1.995(1) | 0.014(1)  |
| 67.4 | 1.996(1) | 0.0140(1) |

## 3.3. Raman spectroscopy

The Raman spectra of the Zn-Ir-O films are shown in **Figure 5**. For the pure ZnO$_x$ films, the spectra contain the characteristic *w*-ZnO vibration bands – $A_1^{LO}$, $E_2^{high}$, and $E_2^{low}$. The bands disappear with the addition of Ir, although some of them are still noticeable for the heated films up to 3.2 Ir at.%.

After the Ir incorporation into the film structure, a wide but intense band appears around 720 cm$^{-1}$, which was for the first time detected in Ref. [14]. It is still unclear what kind of vibrations cause the band; however, it might be attributed to a peroxide ion ($O_2^{2-}$) stretching with a proper O-O distance.[29] The 720 cm$^{-1}$ band begins to overlap with a wide band formed at lower frequencies for the non-heated film with 29.4 Ir at.%. At 44.6 Ir at.% the 720 cm$^{-1}$ band is completely blurred and a wide band has been formed in the range from 300 to 700 cm$^{-1}$. The spectrum remains unchanged up to the pure $a$-IrO$_{2-x}$ film without appearance of additional bands.

For the heated films, the 720 cm$^{-1}$ band is well detectable up to 36.9 Ir at.%. A wide vibration band around 545 cm$^{-1}$, which can be attributed to the vibration mode E$_g$ of $r$-IrO$_2$, is noticeable in the spectrum for the film with 61.5 Ir at.%. It can be concluded that the heated Zn-Ir-O films with the Ir concentration above 61.5 at.% contain IrO$_2$ nanocrystallites. The spectrum of the pure IrO$_{2-x}$ film contains the vibration band E$_g$ and the overlapped bands B$_{2g}$ and A$_{1g}$. The vibration bands are wider compared to the spectrum of polycrystalline $r$-IrO$_2$ indicating the lower degree of crystallinity.

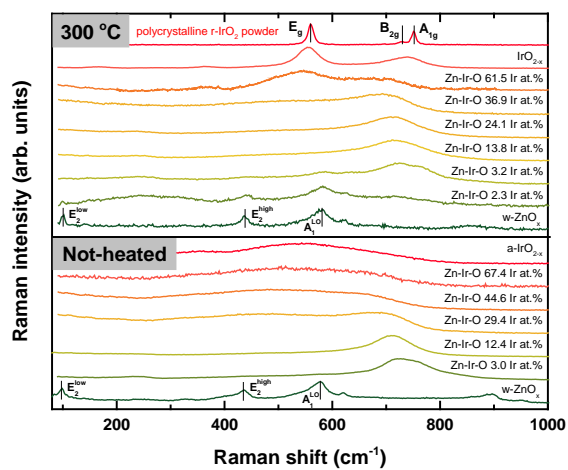

**Figure 5.** Raman spectra of the Zn-Ir-O, $ZnO_x$, and $IrO_{2-x}$ films and the reference compound – polycrystalline bulk $r$-$IrO_2$ taken from Ref. [20].

**3.4. Electrical properties**

The conductivity of the films was measured at a constant temperature of 300 K. The resistivity of the films with low Ir concentration (< 12.4 at.% for the non-heated samples and < 8.0 at.% for the heated samples) is extremely large and exceeds the measurable range. The goal of the Zn-Ir-O deposition was to achieve transparent *p*-type conducting thin films. A relatively high oxygen partial pressure was used in the deposition process to prevent the formation of zinc interstitials, which are donor type defects in the ZnO structure and the possible source of spontaneous *n*-type conductivity. Apparently, the ZnO doping with Ir does not create appropriate defects in the film's structure to sufficiently elevate the conductivity. However, the resistivity in the measurable range decreases exponentially with the Ir concentration (**Figure 6**). The resistivity of the heated films seems to be slightly lower compared to that of the non-heated films.

The Hall effect measurements, except for the heated $IrO_{2-x}$ film, could not be performed, because it was not possible to accurately detect the Hall voltage. For the heated $IrO_{2-x}$ film, the Hall effect measurement shows that it is a *p*-type conductor with the hole concentration of $4.8 \times 10^{22}$ cm$^{-3}$ and the mobility of 0.5 cm$^2$V$^{-1}$s$^{-1}$. The relatively low hole mobility in the $IrO_{2-x}$ film suggests that it might be even lower for the Zn-Ir-O films and could explain why the Hall effect measurements were unsuccessful.

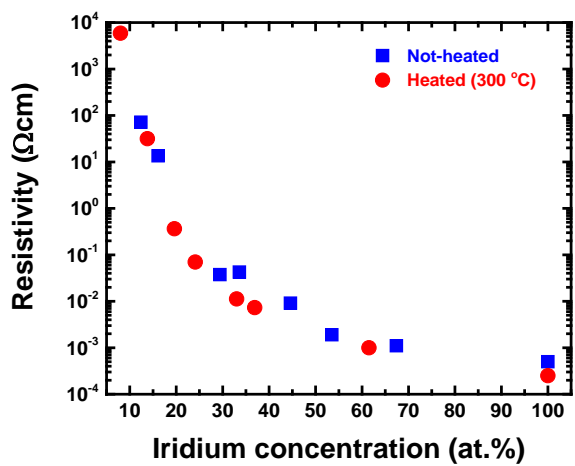

**Figure 6.** Resistivity of the deposited films as a function of Ir concentration.

Thermoelectric measurements were performed to determine the conductivity type of the Zn-Ir-O films. The Seebeck voltage was measured by varying temperature difference from -5 to 5 K, and the Seebeck coefficients were calculated from the slopes of the obtained linear relations. The Seebeck coefficients of the films are plotted in **Figure 7**. Regardless of whether the films were heated or not during the deposition process, a transition from *n*- to *p*-type conductivity was observed upon an increase in the Ir concentration. The films are *n*-type conductors below ≈13 Ir at.%. Above this value all the films are *p*-type conductors.

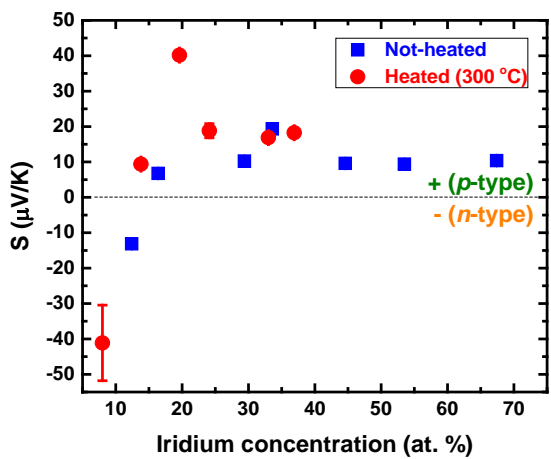

**Figure 7.** Seebeck coefficients of the deposited Zn-Ir-O films as a function of iridium concentration. The error margins of measured values are denoted by error bars or inside the symbols.

The transition suggests that there are several competing conductivity mechanisms and sources of charge carriers in Zn-Ir-O. ZnO is known from its tendency to exhibit spontaneous *n*-type semi-conductivity. Moreover, when properly doped, it can be transformed into a material with metallic conductivity and high visible light transmission.[30,31] Despite the reports that *p*-type doped ZnO has been produced in some experimental works [32], theoretical studies conclude that it is almost impossible to achieve *p*-type conductivity in ZnO.[33] Even if holes are formed in ZnO, they are quenched by charge compensating ionic defects. From our XRD and XAS measurements, it is reasonable to assume that below ≈13 Ir at.% the nano-crystalline *w*-ZnO phase is still present in the Zn-Ir-O films, which would be favourable for the *n*-type conductivity. If the Ir concentration is increased, the structure of the films becomes amorphous. At least for the heated films, the Raman spectroscopy results show an existing $IrO_2$ phase in the Zn-Ir-O films already at ≈60 Ir at.%. Pure $IrO_2$ films are *p*-type conductors, and this fact is supported by the Hall effect measurement of the heated $IrO_{2-x}$ film in this study.

## 4. Conclusion

In this study the structure and the electrical properties of zinc-iridium oxide (Zn-Ir-O) thin films with various iridium concentrations were investigated. The films were deposited by reactive DC magnetron co-sputtering. The Zn-Ir-O structure changes from nano-crystalline to amorphous with the Ir concentration as confirmed by both XAS and XRD measurements; however, the nano-crystalline phase is preserved in a wider Ir concentration range if the substrates are heated during the deposition. XAS data also show that for amorphous films further increase in the Ir content leads to more disordered structures around Zn ions. Tetrahedral coordination of Zn ions was identified by XANES analysis and confirmed by RMC simulations of the Zn K-edge EXAFS spectra as the predominant complex. It was found that the heated film structure contains *r*-IrO$_2$ nanocrytallites above ≈60 Ir at.%. After Ir incorporation into the film structure, the intense Raman band appears at 720 cm$^{-1}$, which then is blurred upon an increase in the Ir concentration. The transition from *n*- to *p*-type conductivity was observed, when the Ir concentration was increased up to ≈13 Ir at.% for both non-heated and heated films. Above 13 Ir at.% all the films are *p*-type conductors. The thermoelectric measurements together with the Hall effect measurements of pure IrO$_{2-x}$ show that the *p*-type conductivity in Zn-Ir-O films is stable at least up to 300 °C and in the wide range of Ir content (13–100 at.% Ir).

**Acknowledgements**

We greatly acknowledge the financial support via ERDF Project No. 1.1.1.1/18/A/073. The authors are greatly indebted to prof. Anders Hallén (KTH) and prof. Mattias Hammar (KTH) for many stimulating discussions.

Received: ((will be filled in by the editorial staff))
Revised: ((will be filled in by the editorial staff))
Published online: ((will be filled in by the editorial staff))

((**For Reviews and Perspectives,** please insert author biographies and photographs here for those authors who should be highlighted, max. 100 words each))

Author Photograph(s) ((40 mm broad, 50 mm high, color or grayscale))

The table of contents entry should be 50–60 words long and should be written in the present tense. The text should be different from the abstract text.

C. Author 2, D. E. F. Author 3, A. B. Corresponding Author* ((same order as byline))

**Title** ((no stars))

ToC figure ((Please choose one size: 55 mm broad × 50 mm high **or** 110 mm broad × 20 mm high. Please do not use any other dimensions))

**Amorphous *p*-type conducting Zn-*x*Ir oxide (*x*>0.13) thin films deposited by reactive magnetron co-sputtering**

*Martins Zubkins, Janis Timoshenko, Jevgenijs Gabrusenoks, Kaspars Pudzs, Andris Azens, Qin Wang, Juris Purans\**

Dr. M. Zubkins, Dr. J. Timoshenko, Dr. J. Gabrusenoks, Dr. K. Pudzs, Dr. A. Azens, Dr. J. Purans
Institute of Solid State Physics, University of Latvia, Kengaraga 8, LV-1063 Riga, Latvia
E-mail: purans@cfi.lu.lv

Dr. Q. Wang
*Department of Smart Hardware, Reasearch Institutes of Sweden, box 1070, 164 25 Stockholm, Sweden*

Zinc-Iridium Oxide (Zn-Ir-O) thin films have been demonstrated as a *p*-type conducting material. However, the origin of *p*-type conductivity is still unclear. This research focuses on the structure and the electrical properties of Zn-Ir-O films in the large Ir concentration range. The films are deposited by reactive DC magnetron co-sputtering. Additionally, two different substrate temperatures are used – without intentional heating and 300 °C. The structure of the films becomes X-ray amorphous by increasing the Ir concentration, and structural disorder increases upon further increase in Ir content. The nano-crystalline *w*-ZnO structure remains in the wider Ir concentration range if a substrate is heated. At low Ir concentration range, the transition from *n*- to *p*-type conductivity is observed regardless whether the substrates are heated or not. Electrical resistivity decreases exponentially with the Ir concentration in the Zn-Ir-O films and it is slightly lower in the case of the heated films.

## 1. Introduction

In order to develop transparent electronics, it is still essential to search for reliable transparent *p*-type semiconductors. Kawazoe et al.[1] reported the *p*-type conductivity in the highly transparent thin film of copper aluminium oxide. Since this publication, a variety of *p*-type transparent conductive oxides (*p*-TCOs) have been investigated. Despite these results, their insufficient transparency and low conductivity continue to restrict the technological applications of the *p*-TCOs.[2-3] Several precious metal-based transparent oxides have also shown the *p*-type conductivity, including polycrystalline Zn$M_2$O$_4$ (*M*= Ir, Rh. Co) films.[4-6]

It has been shown that the conductivity type of zinc-iridium oxide (Zn-Ir-O) films changes from *n*- to *p*-type by increasing an Ir concentration from 12 to 16 at.%.[7] However, a conductivity mechanism and the origin of the transition are unknown. According to DFT calculations performed in Ref.[8] IrO$_x$ ($x$ = 4,5,6) complexes show *p*-type conductivity with the Fermi level up to 0.8 eV from the valance band maximum (VBM) if the Ir concentration is greater than 12.5%. The theoretical calculations of the substitution defects of Ir$^{2+}_{Zn}$ and Ir$^{3+}_{Zn}$ in Ir-doped ZnO predict localized energy states in the band gap which would reduce the transmittance of ZnO:Ir films.[9,10] The visible light transmittance decrease in Zn-Ir-O with Ir concentration has been observed experimentally.[11,12] The structural analysis of doped IrO$_2$ electrocatalysts with the general composition of Ir$_{1-x}M_x$O$_2$ (*M* = Co, Ni, and Zn, $0.05 \leq x \leq 0.2$) prepared by a hydrolysis method shows that the doping elements enter the lattice positions in rutile structure of iridium dioxide.[13] Analysis of the local structure of the catalysts based on EXAFS shows that the dopant cations are not homogeneously distributed but have a tendency to form clusters. Zn-Ir-O has hardly ever been studied in a wide Ir concentration range. In this paper, XRD, XAS and Raman techniques are used to investigate the structure of Zn-Ir-O films deposited by reactive DC magnetron co-sputtering. The electrical conductivity as well as the thermoelectric measurements are also presented to determine the electrical conductivity and transition from *n*-type to *p*-type of the films.

**2. Experimental details**

Zn-Ir-O as well as pure $ZnO_x$ and $IrO_{2-x}$ thin films were deposited on soda-lime glass, Ti and polyimide type substrates by reactive DC magnetron co-sputtering in an Ar (20 sccm) + $O_2$ (10 sccm) atmosphere (10 mTorr working pressure). Two types of sputtering methods were used to deposit the studied films in this paper. Part of the films (mainly at high Ir concentration) were deposited by sputtering a metallic Zn (99.95 wt%) target with Ir (99.6 wt%) pieces on the target erosion zone. An Ir concentration was varied by its amount on the Zn target erosion zone. The second part of the samples (mainly at low Ir concentration) were deposited by sputtering metallic Zn (99.95 wt%) and Ir (99.6 wt%) targets simultaneously. The power on the Ir target was used as a composition control parameter. A detailed experimental procedure of both methods can be found in Ref. [7] and [14]. An elemental analysis of the films was carried out by an X-ray fluorescence spectrometer (XRF), Eagle III. Due to the fact that with XRF it is difficult to quantify elements lighter than sodium, our measurements include only the Ir to Zn atomic concentration ratio. To determine the influence of a substrate temperature on the films structure and properties two sets of samples were deposited: one set without intentional substrate heating during the deposition and the second one with the additional heating at a temperature of 300 °C. All the studied samples in this paper together with the deposition parameters are summarized in **Table 1**.

**Table 1.** Deposition parameters, thickness and Ir/Zn atomic concentration ratio of the studied Zn-Ir-O films on glass, Ti, and polyimide substrates.

| Sample | Sputtering target(s) | Sputtering power (W) | Ir area on the Zn target erosion zone (%) | Thickness (nm) | Substrate temperature (°C) | Ir/Zn atomic concentration ratio (%) |
|---|---|---|---|---|---|---|
| $ZnO_x$ | Zn | 200 | - | 388 | Not heated | 0.0 |
| Zn-Ir-O | Zn and Ir | 200 (Zn), 6 (Ir) | - | 566 | Not heated | 1.7 |
| Zn-Ir-O | Zn and Ir | 200 (Zn), 10 (Ir) | - | 581 | Not heated | 3.0 |
| Zn-Ir-O | Zn and Ir | 200 (Zn), 20 (Ir) | - | 659 | Not heated | 5.1 |
| Zn-Ir-O | Ir pieces on Zn | 100 | ≈0.7 | 393 | Not heated | 7.0 |
| Zn-Ir-O | Zn and Ir | 200 (Zn), 40 (Ir) | - | 752 | Not heated | 12.4 |
| Zn-Ir-O | Zn and Ir | 200 (Zn), 70 (Ir) | - | 778 | Not heated | 16.1 |
| Zn-Ir-O | Ir pieces on Zn | 100 | ≈3.0 | 308 | Not heated | 29.4 |
| Zn-Ir-O | Ir pieces on Zn | 100 | ≈5.8 | 244 | Not heated | 33.6 |
| Zn-Ir-O | Ir pieces on Zn | 100 | ≈9.6 | 141 | Not heated | 44.6 |
| Zn-Ir-O | Ir pieces on Zn | 100 | ≈12.2 | 121 | Not heated | 53.5 |
| Zn-Ir-O | Ir pieces on Zn | 100 | ≈14.6 | 95 | Not heated | 67.4 |
| $IrO_{2-x}$ | Ir | 100 | - | 108 | Not heated | 100.0 |
| $ZnO_x$ | Zn | 200 | - | 388 | 300 | 0.0 |
| Zn-Ir-O | Zn and Ir | 200 (Zn), 6 (Ir) | - | 165 | 300 | 2.3 |
| Zn-Ir-O | Zn and Ir | 200 (Zn), 10 (Ir) | - | 164 | 300 | 3.2 |
| Zn-Ir-O | Zn and Ir | 200 (Zn), 15 (Ir) | - | 193 | 300 | 5.6 |
| Zn-Ir-O | Zn and Ir | 200 (Zn), 25 (Ir) | - | 233 | 300 | 8.0 |
| Zn-Ir-O | Zn and Ir | 200 (Zn), 40 (Ir) | - | 190 | 300 | 13.8 |
| Zn-Ir-O | Zn and Ir | 200 (Zn), 70 (Ir) | - | 282 | 300 | 19.6 |
| Zn-Ir-O | Zn and Ir | 200 (Zn), 90 (Ir) | - | 226 | 300 | 24.1 |
| Zn-Ir-O | Zn and Ir | 200 (Zn), 110 (Ir) | - | 286 | 300 | 33.0 |
| Zn-Ir-O | Zn and Ir | 200 (Zn), 130 (Ir) | - | 281 | 300 | 36.9 |
| Zn-Ir-O | Ir pieces on Zn | 100 | ≈14.6 | 185 | 300 | 61.5 |
| $IrO_{2-x}$ | Ir | 100 | - | 176 | 300 | 100.0 |

The XRD measurements were done on a PANalytical X'Pert PRO diffractometer equipped with the Cu anode X-ray tube and a multichannel solid-state detector. The Zn K-edge (9659 eV) and Ir $L_3$-edge (11215 eV) X-ray absorption spectra (XAS) were measured in transmission mode at the SOLEIL synchrotron bending-magnet beamline Samba [15] in ambient conditions. More details of the experiment can be found in Ref. [7]. Raman spectroscopy measurements were performed at room temperature by a SPEX1403 monochromator with multichannel detectors and an inVia Renishaw Raman microscope. Both an Ar laser (514.5 nm) and YAG second harmonics laser (532 nm) were used as the excitation source. The electrical properties of the films were investigated by Hall effect measurement system HMS5000 at room temperature. To determine the sign of the Seebeck coefficient the thermoelectric measurements in plane were performed by a self-assembled measurement system. More details of the system can be found in Ref.[7].

## 3. Results and discussion

### 3.1. XRD measurements

The X-ray diffractograms of the Zn-Ir-O films are shown in **Figure 1**. It is clearly seen that the crystallinity of the films deteriorates when an Ir concentration is increased. Pure $ZnO_x$ films contain nano-crystallites with the structure of wurtzite type ZnO (*w*-ZnO). This structure is observed even for the not-heated sample due to the rapid crystallisation of ZnO at room temperature. An X-ray amorphous structure of ZnO can be obtain if deposition is performed at cryogenic temperatures.[16] The nano-crystallites have a preferred (002) orientation in the direction of *c*-axis. The preferred orientation decreases with the Ir concentration and an additional (101) maximum appears. At this point, both (002) and (101) diffraction maximums are shifted towards lower angles compared to the *w*-ZnO reference X-ray diffractogram (PDF card No.: 01-070-8072). The further increase of the Ir concentration changes the structure from nano-crystalline to X-ray amorphous. The nano-crystalline structure remains in the larger Ir concentration range if the substrate temperature is 300 ºC during the deposition. It is previously shown that the Zn-Ir-O films become completely amorphous in the Ir concentration range from 7 to 16 at.% if the substrates are not heated intentionally.[7] The X-ray amorphous structure remains up to the pure $IrO_{2-x}$ film for the not-heated samples. Two shifted diffraction maximums, which correspond to the (110) and (200) planes of rutile $IrO_2$ (*r*-$IrO_2$) structure (PDF card No.: 00-015-0870), appear for the pure $IrO_{2-x}$ film, which was deposited on the heated substrate.

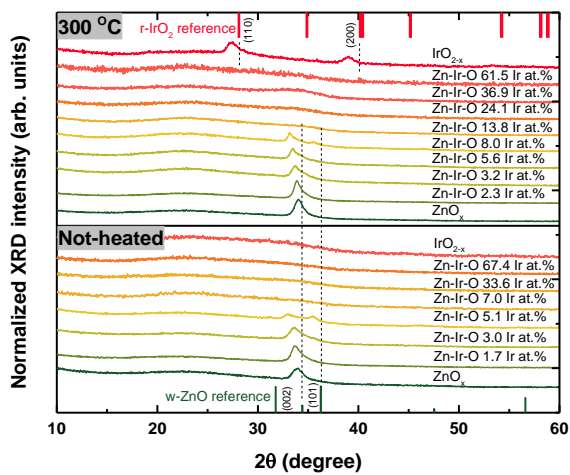

**Figure 1.** X-ray diffractograms of the Zn-Ir-O films at different Ir concentrations for both heated (300 °C) and not-heated samples during the deposition.

### 3.2. XAS measurements

X-ray absorption spectroscopy (XAS) is the one of the most important techniques for investigations of the local atomic structure, and oxidation state of the atom of interest in a broad range of materials, including the X-ray amorphous structures. In this work, we have analysed both the X-ray absorption near-edge structure (XANES) and extended X-ray absorption fine structure (EXAFS) parts of XAS spectra, collected at Ir $L_3$-edge and Zn K-edge. Here XANES spectra, extending up to ca. 40 eV above the absorption edge, are very sensitive to the charge density (i.e., the evolution of the oxidation sate) and bonding motifs around the absorbing atom. EXAFS spectra, in turn, extending from ca. 40 eV up to 1 keV above the absorption edge, are important for the determination of distribution of neighbouring atoms.[17-20]

Zn K-edge XANES spectra (**Figure 2(a)**) for all Zn-Ir-O samples align well with the reference spectrum for bulk *w*-ZnO, suggesting 2+ oxidation state for Zn in Zn-Ir-O films, and the presence of $ZnO_4$ tetrahedra as the main structural units. However, XANES spectra for our thin films appear relatively featureless with respect to that of bulk ZnO, suggesting strongly disordered environment for Zn species. The changes in the relative intensity of the main Zn K-edge XANES features upon an increase in Ir concentration reflect the different degree of distortion of $ZnO_4$ tetrahedra.[21]

Collected Ir $L_3$-edge XANES spectra (**Figure 2(b)**) for Zn-Ir-O films show systematic shift of the main feature toward higher energies upon increased Ir concentration, which can be linked to the changes in the density of available vacant states in the *d*-orbital for the excited $2p_{1/2}$ electrons, and suggests effective increase of Ir oxidation state.[19, 22, 23]

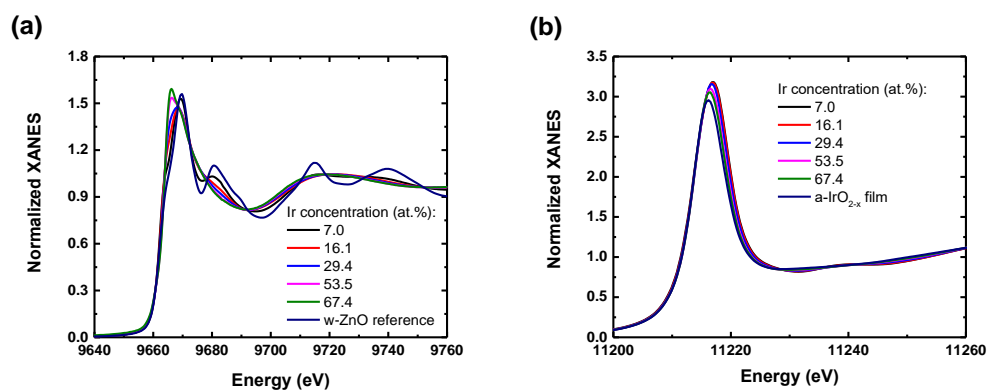

**Figure 2.** Zn K-edge (a) un Ir $L_3$-edge (b) XANES spectra of the Zn-Ir-O films with different Ir atomic concentration and reference compounds – polycrystalline bulk *w*-ZnO taken from Ref. [24] and the amorphous (*a*-) $IrO_{2-x}$ film deposited in this study.

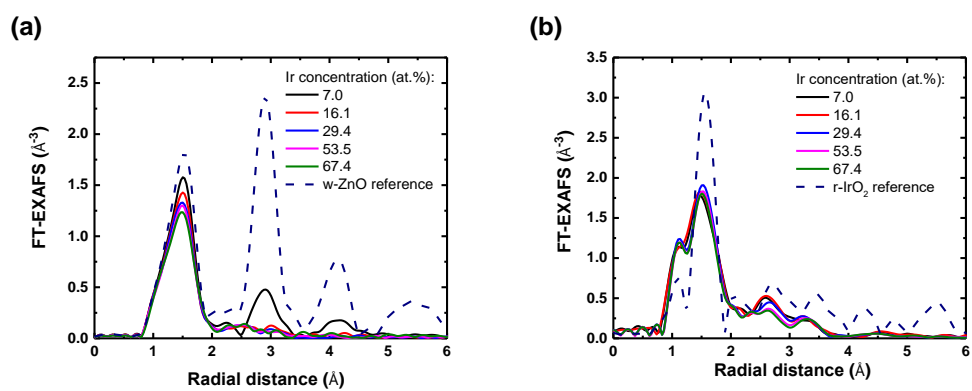

**Figure 3.** Modulus of the FT-EXAFS spectra at Zn K-edge (a) and Ir $L_3$-edge (b) for the Zn-Ir-O films and reference compounds – polycrystalline bulk *w*-ZnO taken from Ref. [24] and *r*-$IrO_2$ taken from Ref. [20]. (Note: distances in the FT-EXAFS do not correspond to the real distances, due to the phase shifts of the signal.)

Further insight into the local structure of Zn and Ir species is provided by the analysis of EXAFS spectra, shown in **Figure 3(a)** and **3(b)**, respectively. In the case of Zn K-edge EXAFS, Fourier-transformed (FT) spectra are dominated by the contribution of first coordination shell (Zn-O bonds). Similar position and the intensity of the main FT-EXAFS peak in all Zn-Ir-O samples confirm that 4-coordinated Zn species with Zn-O bond length similar to that in *w*-ZnO are the main structural units in our thin films, in agreement with XANES results. The slight decrease in the intensity of FT-EXAFS peaks upon an increase in Ir concentration suggests increase in structural disorder (Debye-Waller factors).

Strong structural disorder results also in suppression of more distant FT-EXAFS peaks. Figure 3(a) highlights the fundamental difference between sample with higher Ir content and that with 7.0 at.% Ir. Only in the sample with the lowest Ir concentration we observed presence of significant 2nd and 3rd peaks in FT-EXAFS, which resemble those in *w*-ZnO, suggesting that wurtzite-type structure is locally preserved in this sample. No such peaks were observed for Zn-Ir-O peaks with higher Ir loading, suggesting their amorphous nature. We emphasize here also the lack of significant Zn-Ir bond contributions in our EXAFS data.

To fit the contributions of distant coordination shell, we employ reverse Monte Carlo (RMC) method, as described in Ref. [25] and implemented within EvAX code.[26] Here the EXAFS spectra are fitted in an iterative stochastic process, where we start with ideal *w*-ZnO structure, and introduce small random displacements in the positions of all atoms within 4×4×4 ZnO supercell with periodic boundary conditions, until a good agreement is obtained between experimental Zn K-edge EXAFS spectrum and simulated spectrum for the structure model. For EXAFS spectra calculations we use FEFF8.5 code [27] and include multiple-scattering effects with up to 7 scattering events included. See Ref. [25] for more details. The advantage of RMC approach in comparison to conventional EXAFS fitting is that it allows fitting of contributions of distant coordination shells, but also, crucially, allows fitting of EXAFS spectra for strongly distorted structures, where the conventional EXAFS fitting, which commonly relies on the assumption that the bond length distribution has near-Gaussian shape, results in significant systematic errors.[28]

The results of RMC fitting for Zn K-edge EXAFS spectra are shown in **Figure 4(a)** and **4(b)**. One needs to emphasize here that due to limited information available from EXAFS spectra of our disordered thin films, the obtained structure model is not unambiguous (especially for Ir-rich thin films, where only the first coordination shell contribute significantly to the experimental data). Nevertheless, the good agreement between simulated and experimental EXAFS spectra confirms that the $ZnO_4$ tetrahedral units are the main Zn species in all the samples, and we expect also that our simulations provide reliable information on the shape of the radial distribution function (RDF) at least for the first coordination shell (**Figure 4(c)**).

The obtained RDFs show increasingly asymmetric and broadened shape for Zn-O RDF upon increase in the Ir concentration. The is reflected also in the increased average Zn-O interatomic distance and increased disorder factor, calculated from the atomic coordinates in the final RMC model (**Table 2**).

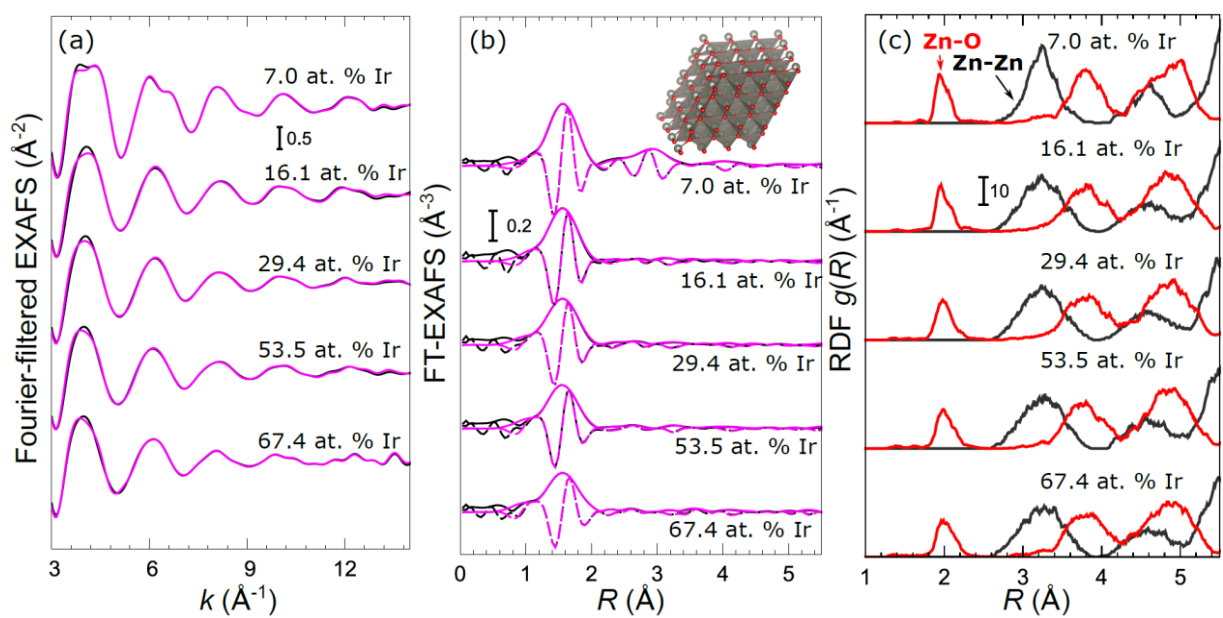

**Figure 4.** RMC fitting for Zn K-edge EXAFS for Zn-Ir-O films with different Ir content. Comparison of experimental spectra with the ones calculated for the final RMC model in *k*-space after Fourier-filtering (a) and in *R*-space after Fourier transformation (b). The final structure model for the sample with 7.0 at.% Ir is shown in the inset. Partial RDFs for Zn-O and Zn-Zn bonds (c). Spectra and RDFs are shifted vertically for clarity.

**Table 2.** Average Zn-O interatomic distances and corresponding disorder factor (Debye-Waller factor $\sigma^2$) for the Zn-Ir-O films as obtained from RMC-EXAFS analysis and Zn K-edge. Uncertainties of the last digit are given in parentheses.

| Ir concentration (at.%) | $R$ (Å) | $\sigma^2$ (Å$^2$) |
|---|---|---|
| 7.0 | 1.991(1) | 0.0081(1) |
| 16.1 | 1.995(1) | 0.0099(2) |
| 29.4 | 1.995(1) | 0.0119(2) |
| 53.5 | 1.995(1) | 0.014(1) |
| 67.4 | 1.996(1) | 0.0140(1) |

### 3.3. Raman spectroscopy

The Raman spectra of the Zn-Ir-O films are shown in **Figure 5**. For the pure ZnO$_x$ films, the spectra contain the characteristic *w*-ZnO vibration bands – A$_1^{LO}$, E$_2^{high}$, and E$_2^{low}$. The bands disappear with the addition of Ir, although some of them are still noticeable for the heated films up to 3.2 Ir at.%.

After the Ir incorporation into the film structure, a wide but intense band appears around 720 cm$^{-1}$, which was for the first time detected in Ref. [14]. It is still unclear what kind of vibrations cause the band; however, it might be attributed to a peroxide ion (O$_2^{2-}$) stretching with a proper O-O distance.[29] The 720 cm$^{-1}$ band begins to overlap with a wide band formed at lower frequencies for the not-heated film with 29.4 Ir at.%. At 44.6 Ir at.% the 720 cm$^{-1}$ band is completely blurred and a wide band has been formed in the range from 300 to 700 cm$^{-1}$. This spectrum remains unchanged up to the pure *a*-IrO$_{2-x}$ film without additional bands.

For the heated films, the 720 cm$^{-1}$ band is well detectable up to 36.9 Ir at.%. A wide vibration band around 545 cm$^{-1}$, which can be attributed to the vibration mode E$_g$ of *r*-IrO$_2$, is noticeable in the spectrum for the film with 61.5 Ir at.%. It can be concluded that the heated Zn-Ir-O films with the Ir concentration above 61.5 at.% contain IrO$_2$ nanocrystallites. The spectrum of the pure IrO$_{2-x}$ film contains the vibration band E$_g$ and the overlapped bands B$_{2g}$ and A$_{1g}$. The vibration bands are wider compared to the spectrum of polycrystalline *r*-IrO$_2$ indicating the lower degree of crystallinity.

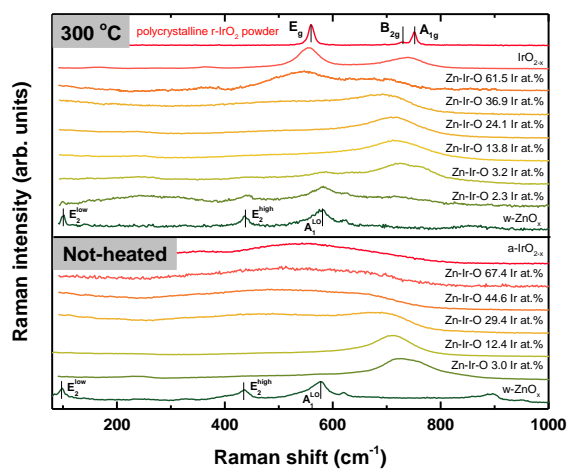

**Figure 5.** Raman spectra of the Zn-Ir-O, ZnO$_x$, and IrO$_{2-x}$ films and the reference compound – polycrystalline bulk *r*-IrO$_2$ taken from Ref. [20].

### 3.4. Electrical properties

The conductivity of the films was measured at a constant temperature of 300 K. The resistivity of the films with low Ir concentration (< 12.4 at.% for the not-heated samples and < 8.0 at.% for the heated samples) is extremely large and exceeds the measurable range. The goal of the Zn-Ir-O deposition was to achieve transparent *p*-type conducting thin films. A relatively high oxygen partial pressure was used in the deposition process to prevent the formation of zinc interstitials, which are donor type defects in the ZnO structure and the possible source of spontaneous *n*-type conductivity. Apparently, the ZnO doping with Ir does not create appropriate defects in the films structure to sufficiently elevate the conductivity. However, the resistivity in the measurable range decreases exponentially with the Ir concentration (**Figure 6**). The resistivity of the heated films seems to be slightly lower compared to the not-heated films. The Hall effect measurements, except for the heated $IrO_{2-x}$ film, could not be performed, because it was not possible to accurately detect the Hall voltage. For the heated $IrO_{2-x}$ film, the Hall effect measurement shows that it is a *p*-type conductor with the hole concentration of $4.8 \times 10^{22}$ cm$^{-3}$ and the mobility of 0.5 cm$^2$V$^{-1}$s$^{-1}$. The relatively low hole mobility in the $IrO_{2-x}$ film suggests that it might be even lower for the Zn-Ir-O films and the reason why the Hall effect measurements were unsuccessful.

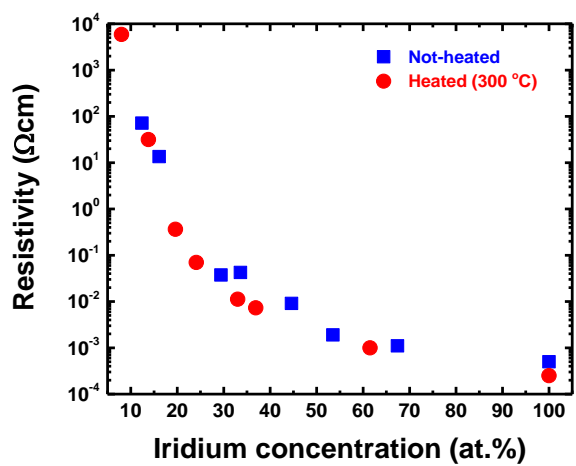

**Figure 6.** Resistivity of the deposited films as a function of Ir concentration.

Thermoelectric measurements were performed to determine the conductivity type of the Zn-Ir-O films. The Seebeck voltage was measured by varying temperature difference from -5 to 5 K, and the Seebeck coefficients were calculated from the slopes of the obtained linear relations. The Seebeck coefficients of the films are plotted in **Figure 7**. Whether the films were heated or not during the deposition process, a transition from *n*- to *p*-type conductivity was observed by increasing the Ir concentration. The films are *n*-type conductors below ≈13 Ir at.%. Above this value all the films are *p*-type.

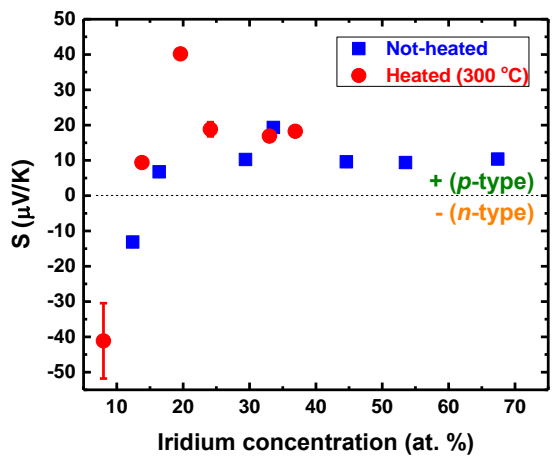

**Figure 7.** Seebeck coefficients of the deposited Zn-Ir-O films as a function of iridium concentration. The error margins of measured values are denoted by error bars or inside the symbols.

The transition suggests that there are several competing conductivity mechanisms and sources of charge carriers in Zn-Ir-O. ZnO is known from its tendency to exhibit spontaneous *n*-type semi-conductivity. Moreover, when properly doped, it can be transformed into a material with metallic conductivity and high visible light transmission.[30,31] Despite that *p*-type doped ZnO has been produced in some experimental works [32], theoretical studies conclude that it is almost impossible to achieve *p*-type conductivity in ZnO.[33] Even if holes are formed in ZnO, they are quenched by charge compensating ionic defects. From the XRD and XAS measurements, it is reasonable to assume that below ≈13 Ir at.% the nano-crystalline *w*-ZnO phase is still present in the Zn-Ir-O films, which would be favourable for the *n*-type conductivity. If the Ir concentration is increased, the structure of the films becomes amorphous. At least for the heated films, the Raman spectroscopy results show an existing $IrO_2$ phase in the Zn-Ir-O films already at ≈60 Ir at.%. Pure $IrO_2$ films are *p*-type conductors, and this fact is supported by the Hall effect measurement of the heated $IrO_{2-x}$ film in this study.

**4. Conclusion**

In this research the structure and the electrical properties of zinc-iridium oxide (Zn-Ir-O) thin films with various iridium concentrations were studied. The films were deposited by reactive DC magnetron co-sputtering. The Zn-Ir-O structure changes from nano-crystalline to amorphous with the Ir concentration as confirmed by both XAS and XRD measurements; however, the nano-crystalline phase remains in the wider Ir concentration range if the substrates are heated during the deposition. XAS data also show that also for amorphous films further increase in the Ir content leads to more disordered structures. It was found out that the heated film structure contains *r*-$IrO_2$ nanocrytallites above ≈60 Ir at.%. After Ir incorporation into the film structure, the intense Raman band appears at 720 cm$^{-1}$, which then blurs with the Ir concentration. The transition from *n*- to *p*-type conductivity was observed, when the Ir concentration was increased up to ≈13 Ir at.% for both not-heated and heated films. Above 13 Ir at.% all the films are *p*-type conductors.

**Acknowledgements**

We greatly acknowledge the financial support via ERDF Project No. 1.1.1.1/18/A/073. The authors are greatly indebted to prof. Anders Hallén (KTH) and prof. Mattias Hammar (KTH) for many stimulating discussions.

Received: ((will be filled in by the editorial staff))  
Revised: ((will be filled in by the editorial staff))  
Published online: ((will be filled in by the editorial staff))

((**For Reviews and Perspectives,** please insert author biographies and photographs here for those authors who should be highlighted, max. 100 words each))

Author Photograph(s) ((40 mm broad, 50 mm high, color or grayscale))

The table of contents entry should be 50–60 words long and should be written in the present tense. The text should be different from the abstract text.

C. Author 2, D. E. F. Author 3, A. B. Corresponding Author* ((same order as byline))

**Title** ((no stars))

ToC figure ((Please choose one size: 55 mm broad × 50 mm high **or** 110 mm broad × 20 mm high. Please do not use any other dimensions))